\documentclass[conference]{IEEEtran}

\usepackage{amsmath,amsfonts,amssymb,epsfig}
\usepackage{cite,footnote,authblk}
\usepackage{color,hyperref}
\hypersetup{
    colorlinks,%
    citecolor=blue,%
    filecolor=blue,%
    linkcolor=blue,%
    urlcolor=blue
}
\definecolor{bluecolor}{rgb}{0,0.,1.}

\definecolor{redcolor}{rgb}{.7,0.,0.}

\newcommand{\pr}[1]{\left( #1\right)}

\newcommand{\es}[1]{\begin{equation}\begin{split}#1\end{split}\end{equation}}

\newcommand{\V}{\mathcal{V}}

\newcommand{\rr}{\mathbf{r}}
\newcommand{\dd}{\mathrm{d}}

\begin{document}

\title{Connectivity in Dense Networks Confined within Right Prisms}
\author[1]{Justin P. Coon}
\author[2,3]{Orestis Georgiou\thanks{orestis.georgiou@toshiba-trel.com}}
\author[3]{Carl P. Dettmann}
\affil[1]{Department of Engineering Science, University of Oxford, Parks Road, Oxford, OX1 3PJ, UK}
\affil[2]{Toshiba Telecommunications Research Laboratory, 32 Queens Square, Bristol, BS1 4ND, UK}
\affil[3]{School of Mathematics, University of Bristol, University Walk, Bristol, BS8 1TW, UK}
\maketitle

\IEEEpeerreviewmaketitle

%\pagestyle{plain}

%%%%%%%%%%%%%%%%%%%%%%
%\thispagestyle{fancy}

\begin{abstract}
We consider the probability that a dense wireless network confined within a
given convex geometry is fully connected. We exploit a recently reported theory to develop a systematic methodology for analytically characterizing the connectivity probability when the network resides within a convex right prism, a polyhedron that accurately models many geometries that can be found in practice. To maximize practicality and applicability, we adopt a general point-to-point link model based on outage probability, and present example analytical and numerical results for a network employing $2 \times 2$ multiple-input multiple-output (MIMO) maximum ratio combining (MRC) link level transmission confined within particular bounding geometries. Furthermore, we provide suggestions for extending the approach detailed herein to more general convex geometries.
\end{abstract}

%%%%%%%%%%%%%%%%%%%%%

\section{Introduction}

Wireless multihop relay networks have received a lot of attention recently due to their
ability to improve coverage and, thus, capacity in a geographical sense. Many
of these networks -- such as mesh, vehicular, wireless
sensor, and \emph{ad hoc} networks -- possess commonality insomuch as
the number and distribution of nodes in the network is often random. A
considerable amount of research on random networks has been conducted in the
past (see, e.g., \cite{Gupta1998,Leveque07,Balister2008,Haenggi2009,Li2009}). From a
communications perspective, it is of great importance to understand the
connectivity properties of such networks since this can lead to
improved protocols  \cite{rajaraman2002topology,bandyopadhyay2003energy}
and network deployment practices~\cite{Ravelomanana2004,romer2004design,younis2008strategies}.

The fundamental question of network connectivity is: what is the probability that all nodes in the
network are connected? The answer is, of
course, related to a number of system properties, such as the fading environment, the
path loss model, and the geometry in which the network
resides. While the first two properties have been thoroughly studied within the construct of network analysis (see, e.g.,~\cite{Bettstetter2004,Bettstetter2005,miorandi2008impact}), the effect that the confining geometry has on connectivity is altogether more complicated to observe analytically.  

Until recently, geometric considerations were limited to simplistic scenarios, including cases where networks are located within circles or squares in two dimensions, or on/in spheres or cubes in three dimensions~\cite{Mao2012,Khalid2013}.  Alternatively, a common, less accurate approach has been to ignore the effects that boundaries play on connectivity altogether, either explicitly or by adopting a network model that implicitly renders such effects negligible (cf. the model used in~\cite{Mao2012}).  Notable progress was made on the topic of geometric effects in~\cite{Coon2012a}, which disclosed a representation for the probability of connectivity for confined random networks that was shown to be accurate in the high density regime.  Moreover, \cite{Coon2012a} also gave a formula for this observable, in which the probability was split into additive components, each corresponding to a particular feature of the bounding geometry.  

Although~\cite{Coon2012a} completely characterized the network connectivity probability in the dense regime, little has been done to apply the framework presented therein to describe scenarios that might be encountered in practice.  To this end, in this paper we develop a method to explicitly analyze the probability that a network contained within a \emph{convex right prism}\footnote{A right prism is a polyhedron with a $q$-sided polygonal base, a copy of this polygon translated in a direction normal to the plane in which the original resides, and $q$ rectangles connecting the respective polygonal sides.  A cube is an example.} is connected.  Right prism bounding geometries are interesting and useful to consider from an engineering perspective since they model many practical scenarios, such as networks confined within a building or room.  
To demonstrate the versatility of our approach, we provide further details through two worked examples, in which we analyze the connectivity probability of a network employing diversity transmission/reception techniques.  
We also validate this analysis with numerical results. 
Finally, we suggest a method that can be used to extend the techniques detailed herein to more general convex domains.

\section{System Model and Background\label{sec:model}}
\subsection{Probability of Connectivity}
Consider a network formed of $N$ randomly distributed nodes with locations $\mathbf{r}_{i}%
\in\mathcal{V}\subseteq\mathbb{R}^{d}$ for $i=1,2,\ldots,N$ according to a
uniform density $\rho=N/V$, where $V= \vert \mathcal{V} \vert $ and
$ \vert \cdot \vert $ denotes the size of the set. Here, we use the
Lebesgue measure of the appropriate dimension $d$. We have that two arbitrary nodes $i$ and $j$ are directly
connected with probability $H(|\mathbf{r}_{i} - \mathbf{r}_{j}|)$, which we
write as $H(\mathbf{r}_{ij})  $, where $|\cdot|$ is the
appropriate distance function. 
In what follows, we are principally concerned with the fundamental concept of whether or not a set of nodes can connect to form a multihop network. 
Consequently, we have neglected the impact of interference by assuming a low traffic load and/or an efficient MAC layer protocol, and thus our model falls within the paradigm of delay tolerant networking for example.

It was shown in~\cite{Coon2012a} that the probability that a randomly deployed network is fully connected can be written as
\begin{equation}
P_{fc} =1-\rho\int_{\mathcal{V}}e^{-\rho\int_{\mathcal{V}}H(  \mathbf{r}%
_{12})  \mathrm{d}\mathbf{r}_{1}}(  1+O(  N^{-1})
)  \mathrm{d}\mathbf{r}_{2}\label{eq:Pfc} 
,
\end{equation}
in the dense regime where $N$ is large.
Note that similar results have been reported elsewhere e.g. in \cite{Mao2012}, \cite{haenggi2008geometric} and \cite{miorandi2008impact}.
Equation (\ref{eq:Pfc}) signifies the asymptotic equivalence of the network's minimum degree distribution and $P_{fc}$ (rigorously proven in \cite{penrose1999k})
and suggests that in the high density limit, \textquotedblleft
hard to connect\textquotedblright\ regions of the available domain
$\mathcal{V}$ govern the probability of connectivity. This is because the outer integral in (\ref{eq:Pfc}) is
dominated where the integral in the exponent is small, which occurs at
corners, edges, and faces in three dimensions. For nodes located near these boundary features, the
volume in range for communication is small.  

Provided the pair connectedness function $H$ decays suitably quickly with increasing $r = |\mathbf r_i - \mathbf r_j|$, this dependence upon specific geometric properties points to a reformulation of~\eqref{eq:Pfc} as a summation of terms, each corresponding to a particular boundary object, which can be written as
\begin{equation}
P_{fc}\approx1-\sum_{\ell=0}^{d}\sum_{k_{\ell}}\rho^{1-\ell}G_{k_{\ell}}V_{k_{\ell}}%
e^{-\rho\omega_{k_{\ell}}\int_{0}^{\infty}r^{d-1}H (  r )  \mathrm{d}%
r}\label{eq:general_formula},
\end{equation}
where $G_{k_{\ell}}$ is a geometrical factor for
each object $k_{\ell}$ of codimension $\ell$ and $V_{k_{\ell}}$ is the corresponding
$d-\ell$ dimensional volume of the object with solid angle $\omega_{k_{\ell}}$.
The interested reader is directed towards \cite{Coon2012a} for further details on scaling properties of equations \eqref{eq:Pfc} and \eqref{eq:general_formula}.

The general formula given in~\eqref{eq:general_formula} is an interesting theoretical result.  However, significant effort is required to apply this theory to the analysis of practical systems.  As a key contribution of this work, in section~\ref{sec:rp} we present a systematic methodology based on the exploitation of~\eqref{eq:Pfc} and~\eqref{eq:general_formula} that can be used to obtain an accurate expression for $P_{fc}$ under the assumption that the network in question is confined within a convex right prism.

\subsection{A Note on Pair Connectedness Functions}
The probability $P_{fc}$ is a functional of $H$, which we assume to be identical for all point-to-point links in the network.  We define $H(\mathbf r_{ij})$ as the complement of the information outage
probability between nodes $i$ and $j$, i.e.,
\begin{equation}
  H(  r)  = P(  \log_{2}(  1+\mathsf{SNR}(r)\cdot X) \geq R_0),
\label{eq:Hij}%
\end{equation}
where $X$ denotes the random variable signifying the normalized power of the channel
between two nodes; $\mathsf{SNR}$ is the signal-to-noise ratio (SNR) at the receiver, which is a function of the distance $r = |\mathbf r_i - \mathbf r_j|$; and $R_0$ is the target error-free transmission rate. It should be noted that other pair connectedness models can
easily be chosen, such as a model based on the average bit-error rate of a
point-to-point link.

The received power decreases with distance like
$r^{-\eta}$ where $\eta$ is the path loss exponent.
Typically, $\eta=2$ if propagation occurs in free space, with $\eta>2$ in
cellular/cluttered environments or through walls~\cite{Paulraj2003}. It follows that the SNR at the
receiver (assuming a fixed transmit power and a sufficiently narrow
bandwidth) also decays like
$r^{-\eta}$. Thus, we can write
\begin{equation}
H(  r )  =1-F_{X} (  \beta r^{\eta} ),
\label{eq:Hij2}%
\end{equation}
where $F_{X}$ is the cumulative distribution function of $X$, and
$\beta$ is a constant, which depends upon the transmission frequency, the receiver noise power, and the transmit power. 
It is not difficult to see that $\beta$ is responsible for the effective communication range $r_0$ defined through the relation $r_0= \beta^{-1/\eta}$. 
Note that in the limit of $\eta\to\infty$, the pair connectedness function $H$ is no longer probabilistic but converges
to the geometric disk model, with an on/off connection range at the limiting $r_0$.
In section~\ref{sec:numerics}, we will elaborate further on pair connectedness functions in the context of specific examples.

\section{Connectivity Analysis for Convex Right Prism Bounding Geometries}
\label{sec:rp}
In this section, we present the main contribution of our work: a general methodology for expressing the probability of connectivity for a network confined to a convex right prism in the form of~\eqref{eq:general_formula}.  We begin with a general approach, then provide an example application.

\subsection{General Approach \label{sec:general}}
To derive $P_{fc}$, we must evaluate the integral (cf.~(\ref{eq:Pfc}))%
\begin{equation}
\int_{\mathcal{V}}e^{-\rho\int_{\mathcal{V}}H (  \mathbf{r}_{12} )
\mathrm{d}\mathbf{r}_{1}}\mathrm{d}\mathbf{r}_{2},
\label{eq:integral_for_prism}%
\end{equation}
for each local feature of the bounding geometry. The general
method that is taken can be outlined as follows for $d=3$ dimensions.  We begin by considering features with the lowest dimension,
i.e., corners. We then consider edges, faces, and
finally the bulk of the prism. At each step, we ignore objects of lower
dimension. It will be observed that this is a particularly powerful
approach when analyzing the effects that the bulk and faces have since the
surface (volume) of a right prism is locally equivalent to that of a sphere of
the appropriate surface area (volume). Once all contributions are calculated, they are added together and multiplied by $\rho$ to obtain the probability of an isolated node at high density (cf.~\eqref{eq:Pfc}).  The complement of this probability is the desired expression for $P_{fc}$.  We now consider the calculation of each contribution in turn.

\subsubsection{Corners}
For a right prism, each corner is formed by the intersection of three planes oriented such that at least one edge connected to the vertex is
normal to an adjoining face. This suggests the problem should be cast in terms of cylindrical
coordinates, with the vertex being the origin and the $z$-axis oriented along the edge that joins the
two identical polygons.

The distance between two points in cylindrical coordinates is given by%
\[
|\mathbf{r}_{1} - \mathbf{r}_{2}|  =\sqrt{r_{1}^{2}+r_{2}%
^{2}-2r_{1}r_{2}\cos (  \theta_{1}-\theta_{2} )  + (  z_{1}%
-z_{2} )  ^{2}}%
,\]
where $ \mathbf r_i = (  r_{i},\theta_{i},z_{i} )  $. In order to evaluate the inner integral of
(\ref{eq:integral_for_prism}) near the corner, we let $\mathbf{r}_{2}$ be
located near the corner and expand $H (  \mathbf{r}_{12} )  $ near
$r_{2}=0$ and $z_{2}=0$ up to first order.  Thus, we can approximate the integral in the exponent as
\es{\label{eq:MH_corner}
  \int_{\mathcal{V}}H (  &\mathbf{r}_{12} )
\mathrm{d}\mathbf{r}_{1} = \int_{\mathcal{V}}r_1 H\bigg(\sqrt{r_1^2 + z_1^2}\bigg) \dd r_1 \dd \theta_1 \dd z_1 \\
 &- \int_{\mathcal{V}}  \frac{r_1 z_1 z_2}{\sqrt{r_1^2 + z_1^2}}H^\prime\bigg(\sqrt{r_1^2 + z_1^2}\bigg) \dd r_1 \dd \theta_1 \dd z_1 \\
 &- \int_{\mathcal{V}} \frac{r_1^2 r_2 \cos(\theta_1 - \theta_2)}{\sqrt{r_1^2 + z_1^2}}H^\prime\bigg(\sqrt{r_1^2 + z_1^2}\bigg)\dd r_1 \dd \theta_1 \dd z_1.
}
The region of integration is $\mathcal V = [0,\infty)\times [0,\vartheta) \times [0,\infty)$
where $\vartheta$ is the dihedral angle of the corner (satisfying $0 < \vartheta < \pi$). Note that
semi-infinite integration is allowed here because $H$ is
exponentially decreasing but the system size is large\footnote{In particular,
if $L$ denotes the typical length of the geometry, then we require
$\sqrt{\beta}L\gg1$ (cf.~\cite{Coon2012a} for more details).}.
Using~\eqref{eq:MH_corner},~\eqref{eq:integral_for_prism} can be evaluated over the same space $\mathcal V$. All that remains is to enumerate the $2q$ corners for a prism constructed from
a $q$-sided polygon.

\subsubsection{Edges}

Now we consider geometric features of dimension one: edges. Let $L$ be the
length of the edge in question. The calculations for this case are also
facilitated by the use of cylindrical coordinates, but where the origin is
located at the center of the edge. Thus, the corners are located at $\pm L/2$
and the angle of the corner is $\vartheta$. Since we wish to ignore effects
from corners, faces, and the bulk, we expand $H$ about $r_{2}=0$ and $z_{2}=0
$ to first order and evaluate the inner integral in~\eqref{eq:integral_for_prism}, which gives~\eqref{eq:MH_corner}, but where $\mathcal V = [0,\infty) \times [0,\vartheta) \times (-L/2,L/2)$.  The outer integral can then be performed over the same space.  Finally, we enumerate the $3q$ edges of the prism.

\subsubsection{Faces}
For the contribution of the faces to the full-connectivity probability, we
employ a \emph{local equivalence} argument that allows us to greatly simplify
the analysis. Specifically, we have already accounted for the corner and edge calculations
above, and thus we ignore contributions from these features when considering
faces. Thus, one can imagine deforming a prism of surface area $S$ into a
sphere of the same surface area, the radius $R$ of which is defined by the
relation $S=4\pi R^{2}$. For a general right prism, the surface area is given
by%
\[
S=2B+ph
\]
where $B$ is the area of the base ({i.e.}, the $q$-sided polygon), $p$ is
the base perimeter, and $h$ is the height. This argument allows us to treat the surface area boundary contribution to $P_{fc}$ of any convex right prism we
wish.

Using spherical coordinates along with the fact that the distance between
nodes at $\mathbf{r}_{1}$ and $\mathbf{r}_{2}$ is given by%
\[
|\mathbf{r}_{1} - \mathbf{r}_{2}|  =\sqrt{r_{1}^{2}+r_{2}%
^{2}-2r_{1}r_{2}\cos\theta} 
\]
where $\rr_i=(r_i,\theta_i,\phi_i)$, $r_{i}= \vert \mathbf{r}_{i} \vert $ and $\theta\in [
0,\pi ]  $ is the angle between the nodes, we expand $H$ near the surface
of the sphere ({i.e.}, $r_{2}=R$) to first order and perform the inner integral in
(\ref{eq:integral_for_prism}) to obtain%
\es{
\label{eq:MH_face}
  \int_{\mathcal{V}}H (  &\mathbf{r}_{12} )
\mathrm{d}\mathbf{r}_{1} = \\
&\int_{\mathcal{V}}r_1^2 \sin \theta H\bigg(\sqrt{R^2 + r_1^2 -2 R r_1 \cos \theta }\bigg) \dd r_1 \dd\theta \dd \phi \\
 &+ \int_{\mathcal{V}}  r_1^2 \sin \theta \frac{(r_2 - R)(R - r_1 \cos \theta)}{\sqrt{R^2 + r_1^2 -2 R r_1 \cos \theta }} \\
 &\times H^\prime\bigg(\sqrt{R^2 + r_1^2 -2 R r_1 \cos \theta }\bigg) \dd r_1 \dd\theta \dd \phi.
}
The region of integration is $\mathcal V = [0,R)\times[0,\pi)\times[0,2\pi)$.  Using this expression, we can evaluate the outer integral in~\eqref{eq:integral_for_prism}, the result of which is a function of the radius $R$.  This analysis can be generalized to any right prism by substituting $S = 4\pi R^2$. The faces do not need to be enumerated since we have accounted
for all faces through the local equivalence argument.

Finally, it is worth mentioning that face contributions can also be calculated
in a lengthy manner by using Cartesian coordinates. This works well for
rectangular sides; however, one must be more careful when considering more
general $q$-sided polygons. In any case, it is straightforward to show for
certain scenarios that the proposed approach to the calculation yields
identical results in the high density regime to the more complex Cartesian approach.

\subsubsection{Bulk}

For the bulk contribution, we apply the same local equivalence argument that
was used for the face contributions, but where the expansion in $H$ is
performed at $r_{2}=0$. In other words, we consider a sphere of radius $R$
determined by the relation $V=\frac{4}{3}\pi R^{3}$, where%
\[
V=Bh
\]
is the volume of the right prism containing the network.

Expanding $H$ about the origin to first order, we can perform the inner
integration to obtain%
\es{
\label{eq:MH_bulk}
  \int_{\mathcal{V}}H (  \mathbf{r}_{12} )
\mathrm{d}\mathbf{r}_{1}& = \int_{\mathcal{V}}r_1^2 \sin \theta H(r_1)\dd r_1 \dd\theta \dd \phi \\
& + \int_{\mathcal{V}}  r_1^2 r_2 \cos \theta \sin \theta H^\prime(r_1)\dd r_1 \dd\theta \dd \phi .
}
where the integration is performed over $\mathbb R^3$ since we assume pair connectedness decays quickly compared to the size of the network domain and the node at $\mathbf r_2$ is located at the origin.  However, we account for the volume of the network domain when evaluating the outer integral by letting $\mathcal V = [0,R)\times[0,\pi)\times[0,2\pi)$ when we perform the calculation.  The volume of the right prism in question can be incorporated in a manner similar to that discussed for the face calculations above.

\subsection{Example 1: the ``House'' Prism\label{sec:numerics}}
\begin{figure}
[ptb]
\begin{center}
\includegraphics[width=2.45in
]%
{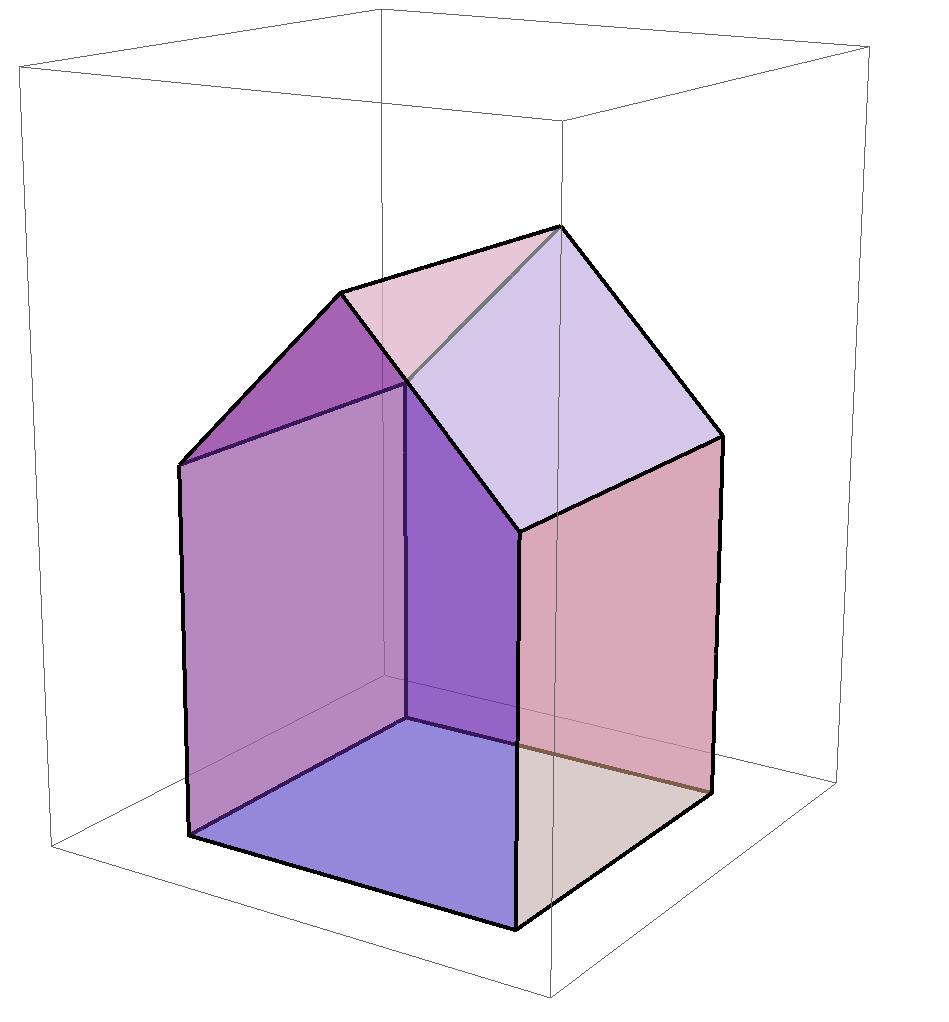}%
\caption{The \textquotedblleft house\textquotedblright\ prism considered for
Example 1. The base is a square of side $L$, the apex is a right
angle, and the total height is $3L/2$.}%
\label{fig:house}%
\end{center}
\end{figure}

It is instructive to consider the approach developed above in the context of a practical example.  
We focus on a network operating in a right prism that resembles a ``house'', as illustrated in Fig.~\ref{fig:house}. The base of
the house is a square of side $L$. The height of the prism is $3L/2$,
and the apex is right angled.

With regard to the pair connectedness function, we wish to demonstrate the versatility of our theory.  Thus, we adopt a slightly more complicated model than the standard scalar Rayleigh fading assumption.  Instead, we model each point-to-point link as a $2 \times 2$ MIMO channel comprised of independent, identically distributed Rayleigh fading constituent channels and assume
beamforming is applied at the transmitter of each node along with maximum ratio
combining (MRC) at the receiver~\cite{Kang2003}. Under the condition that the path loss exponent $\eta = 2$, the pair connectedness function for this
so-called MIMO MRC channel becomes
\begin{equation}\label{eq:H2x2}
H (  r )  =e^{-\beta r^{2}} (  \beta^{2}r^{4}+2-e^{-\beta r^{2}%
} ).
\end{equation}
Note that both simpler and more complicated systems can be studied easily by adjusting the definition of $H$.

First, consider the corner contributions.  Using the expression for $H(r)$ given above, we can evaluate~\eqref{eq:MH_corner} to obtain
\begin{multline}
\int_{\mathcal{V}}H (  \mathbf{r}_{12} )
\mathrm{d}\mathbf{r}_{1}  =  \frac{1}{8\beta}\Bigg(  14z_{2}\vartheta+\frac{23-\sqrt{2}}{2}\sqrt
{\frac{\pi}{\beta}}\vartheta \\
\qquad+7\pi r_{2} (  \sin\theta_{2}-\sin (
\theta_{2}-\vartheta )   )  \Bigg) \label{eq:inner_integral_corner} ,
\end{multline}
where $\vartheta$ is the angle of the corner with $0<\vartheta<\pi$.  Now we may calculate the outer integral of (\ref{eq:integral_for_prism}%
), which gives%
\begin{equation}
\int_{\mathcal{V}}e^{-\rho \int_{\mathcal{V}}H (  \mathbf{r}_{12} )
\mathrm{d}\mathbf{r}_{1}  }%
\mathrm{d}\mathbf{r}_{2}  =\frac{256\beta^{3}\csc\vartheta}{343\pi^{2}\rho^{3}\vartheta}%
e^{-\frac{ (  23-\sqrt{2} )  \sqrt{\pi}\rho\vartheta}{16\beta^{3/2}}%
}.\label{eq:corner} ,
\end{equation}
Recall that $\mathcal V$ is taken to be $[0,\infty)\times [0,\vartheta) \times [0,\infty)$ here.

There are ten corners in this prism, six of which have angle
$\vartheta=\pi/2$ and four of which have angle $\vartheta=3\pi/4$. Thus, the
following two contributions to the general formula for $P_{fc}$ arise from the
corners:%
\begin{equation}
C_{1}=6\frac{512\beta^{3}}{343\pi^{3}\rho^{3}}e^{-\frac{ (  23-\sqrt
{2} )  \rho}{32} (  \frac{\pi}{\beta} )  ^{3/2}}%
\end{equation}
and%
\begin{equation}
C_{2}=4\frac{1024\sqrt{2}\beta^{3}}{1029\pi^{3}\rho^{3}}e^{-\frac{ (
23-\sqrt{2} )  3\rho}{64} (  \frac{\pi}{\beta} )  ^{3/2}}.
\end{equation}

Next, we turn our attention to the edge contributions.  Calculating~\eqref{eq:MH_corner} as discussed above leads to an expression that has
$\exp (  -\beta L^{2}/4 )  $ and $\operatorname{erf} (
L\sqrt{\beta/2} )  $ terms. Assuming that $\sqrt{\beta}L\gg1$, we
can approximate $\exp (  -\beta L^{2}/4 )  \approx0$ and
$\operatorname{erf} (  L\sqrt{\beta/2} )  \approx1$, which allows us to write
\begin{multline}
\int_{\mathcal{V}}H (  \mathbf{r}_{12} )
\mathrm{d}\mathbf{r}_{1}  =\frac{1}{4\beta}\Bigg(  \frac{23-\sqrt
{2}}{2}\sqrt{\frac{\pi}{\beta}}\vartheta \\
+7\pi r_{2} (  \sin\theta_{2}%
-\sin (  \theta_{2}-\vartheta )   )  \Bigg).
\end{multline}
The outer integral in (\ref{eq:integral_for_prism}) can
now be evaluated to obtain%
\begin{equation}
\int_{\mathcal{V}}e^{-\rho \int_{\mathcal{V}}H (  \mathbf{r}_{12} )
\mathrm{d}\mathbf{r}_{1}  }%
\mathrm{d}\mathbf{r}_{2}  = 
\frac{16L\beta^{2}\csc\vartheta}{49\pi
^{2}\rho^{2}}e^{-\frac{ (  23-\sqrt{2} )  \sqrt{\pi}\rho\vartheta
}{8\beta^{3/2}}}.\label{eq:edges}%
\end{equation}

All that remains is to enumerate the fifteen edges.  Thirteen edges are right angled: nine of these have length $L$ while the
remaining four have length $L/\sqrt{2}$. The other two edges have angle
$\vartheta=3\pi/4$ and length $L$. Thus, we can write the following two edge
contributions to the high density expression for $P_{fc}$:%
\begin{equation}
E_{1}=L (  9+2\sqrt{2} )  \frac{16\beta^{2}}{49\pi^{2}\rho^{2}%
}e^{-\frac{ (  23-\sqrt{2} )  \rho}{16} (  \frac{\pi}{\beta
} )  ^{3/2}},
\end{equation}
and%
\begin{equation}
E_{2}=2L\frac{16\sqrt{2}\beta^{2}}{49\pi^{2}\rho^{2}}e^{-\frac{ (
23-\sqrt{2} )  3\rho}{32} (  \frac{\pi}{\beta} )  ^{3/2}}.
\end{equation}

For the face contributions, we can apply the local equivalence argument discussed above and substitute~\eqref{eq:H2x2} into~\eqref{eq:MH_face}.  Evaluating the resulting expression gives
\begin{equation}
\int_{\mathcal{V}}H (  \mathbf{r}_{12} )
\mathrm{d}\mathbf{r}_{1} =\frac{\pi}{4\beta} \left(  \frac{23-\sqrt{2}}{2}\sqrt{\frac{\pi}{\beta}%
}+14 (  R-r_{2} )   \right).
\end{equation}
To arrive at this result, it was assumed that $\sqrt{\beta}R\gg1$, which
allows us to make similar approximations to the error functions of the form
$\operatorname{erf} (  c\sqrt{\beta}R )  $ and exponentials of the
form $\exp (  -c\beta R^{2} )  $ for some constant $c>0$ as was done
for edges. The outer integral can now be evaluated
to yield (to leading order in $R$ and $\rho$)%
\begin{equation}
\int_{\mathcal{V}}e^{-\rho \int_{\mathcal{V}}H (  \mathbf{r}_{12} )
\mathrm{d}\mathbf{r}_{1}  }%
\mathrm{d}\mathbf{r}_{2}  =\frac{8\beta R^{2}%
}{7\rho}e^{-\frac{ (  23-\sqrt{2} )  \pi^{3/2}\rho}{8\beta^{3/2}}}.
\end{equation}
Generalizing this result to any right prism, {i.e.}, substituting $S=4\pi
R^{2}$, gives%
\begin{equation}
\int_{\mathcal{V}}e^{-\rho \int_{\mathcal{V}}H (  \mathbf{r}_{12} )
\mathrm{d}\mathbf{r}_{1}  }%
\mathrm{d}\mathbf{r}_{2}  =\frac{2\beta S}{7\pi\rho}e^{-\frac{ (
23-\sqrt{2} )  \pi^{3/2}\rho}{8\beta^{3/2}}}.\label{eq:surfaces}%
\end{equation}
For our example, the total surface area is%
\begin{equation}
S=\frac{11+2\sqrt{2}}{2}L^{2}.
\end{equation}
Substituting this result
into (\ref{eq:surfaces}) gives the contribution of all of the
faces to $P_{fc}$. We denote this contribution as $F$. 

Lastly, we are left with the bulk contribution.  Using~\eqref{eq:MH_bulk}, we can write
\begin{equation}
\int_{\mathcal{V}}H (  \mathbf{r}_{12} )
\mathrm{d}\mathbf{r}_{1}  = \frac{ (  23-\sqrt{2} )  }{4} \left(  \frac{\pi}{\beta} \right)
^{\frac{3}{2}}.
\end{equation}
It readily follows that
\begin{equation}
\int_{\mathcal{V}}e^{-\rho \int_{\mathcal{V}}H (  \mathbf{r}_{12} )
\mathrm{d}\mathbf{r}_{1}  }%
\mathrm{d}\mathbf{r}_{2} =Ve^{-\frac{ (
23-\sqrt{2} )  \pi^{3/2}\rho}{4\beta^{3/2}}} ,
\label{eq:volume}%
\end{equation}
where $V$ is just the volume of a sphere of radius $R$.  The volume of the prism in question is given by
\begin{equation}
V=\frac{5}{4}L^{3}.
\end{equation}
Substituting into (\ref{eq:volume}) yields the bulk contribution to $P_{fc}$,
which we denote by $U$.

Finally, the general formula is obtained through the summation of all
contributions
\begin{equation}
P_{fc}\approx 1-\rho ( U+F+ E_{1}+E_{2} + C_{1}+C_{2} ) .
\label{final1}
\end{equation}
It can easily be seen that this formula has the same structure as
(\ref{eq:general_formula}). 
To aid the reader, the various contributions to the general
formula are outlined in Table \ref{tab:1}. 
\begin{table}[tbp] \centering
\caption{Contributions of the various geometrical features of the ``house''
prism to the general formula for $P_{fc}$ given by
(\ref{eq:general_formula}).  The angle $\vartheta = \pi / 2$ for type-1
corner/edge contributions, and $\vartheta = 3\pi / 4$ for type-2
contributions.}\label{tab:1}%
\begin{tabular}
[c]{|c||c|c|l|l|}\hline
Formula Parameter & Corners & Edges & Faces & Bulk\\\hline\hline
Volume $V_{j_{i}}$ & $1$ & $L,\,\frac{L}{\sqrt{2}}$ &
\multicolumn{1}{|c|}{$S$} & \multicolumn{1}{|c|}{$V$}\\
Solid Angle $\omega_{j_{i}}$ & $\vartheta$ & $2\vartheta$ &
\multicolumn{1}{|c|}{$2\pi$} & \multicolumn{1}{|c|}{$4\pi$}\\
Geometrical Factor $G_{j_{i}}$ & $\frac{256\beta^{3}\csc\vartheta}%
{343\pi^{2}\vartheta}$ & $\frac{16\beta^{2}\csc\vartheta}{49\pi^{2}}$ &
\multicolumn{1}{|c|}{$\frac{2\beta}{7\pi}$} & \multicolumn{1}{|c|}{$1$%
}\\\hline
\end{tabular}%
\end{table}%

\begin{figure}
[ptb]
\begin{center}
\includegraphics[width=3in]%
{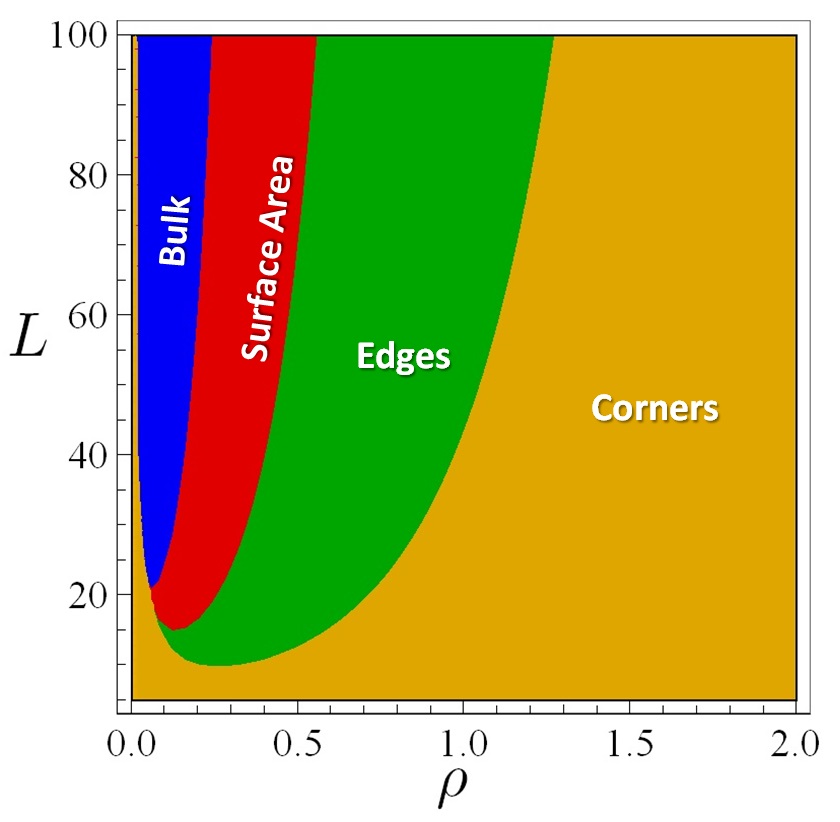}%
\caption{ Parameter space $(\rho,L)$ showing the dominant contribution to the connectivity $P_{fc}$ in the ``house" domain between the bulk $U$ (shown in blue), the surface area $F$ (shown in red), the edges $E_1+E_2$ (shown in green), and the corners $C_1+C_2$ (shown in yellow).}%
\label{fig:phase}%
\end{center}
\end{figure}

It is important to notice that the exponent of each term in \eqref{final1} is smaller by a factor of $2$ when reading from left to right. 
Therefore, at high enough densities $C_2$ (i.e. the sharpest corner) will be the dominant contribution which dictates $P_{fc}$. 
It is however possible that at finite densities, other components dominate over $C_2$. This will depend on the values of $G_{j_{i}} V_{j_{i}}$ controlled by $\beta$ and $L$.
To see this, we set $\beta=1$ and plot in Fig.~\ref{fig:phase} using different colors the regions of the $(\rho,L)$ parameter space for which the bulk $U$ (shown in blue), the surface area $F$ (shown in red), the edges $E_1+E_2$ (shown in green), and the corners $C_1+C_2$ (shown in yellow) dominate the performance of $P_{fc}$.
Fig.~\ref{fig:phase} however should be observed with caution as it is based on asymptotic expansions requiring that $N\gg 1$ and $\sqrt{\beta}L \gg 1$. 
Moreover, to maintain simplicity and tractability in our derivations, expansions were typically only to first order and indeed second-order corrections may be employed to improve accuracy at lower densities.
Effectively this means that the lower left corner of the Fig.~\ref{fig:phase} is not an accurate representation of the parameter space of $P_{fc}$.

\subsection{Numerical Results for Example 1}

\begin{figure}
[ptb]
\begin{center}
\includegraphics[width=3in
]%
{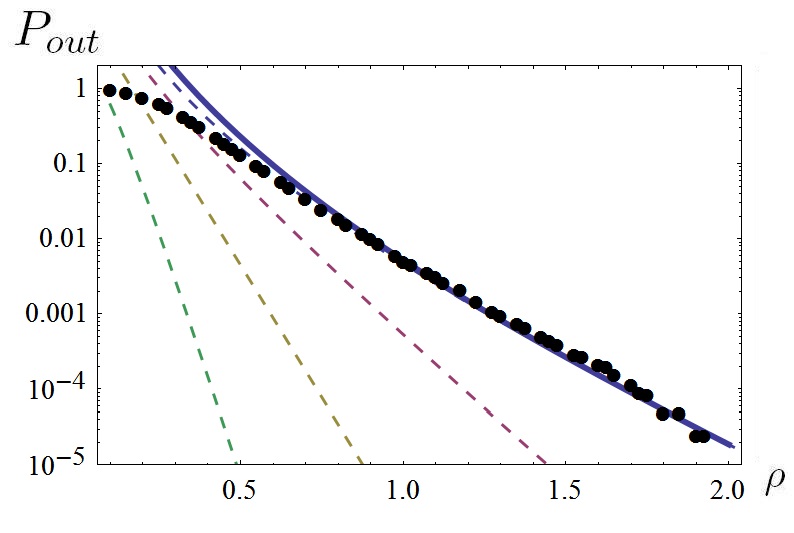}%
\caption{Analytical and numerical results for the network outage probability $P_{out} = 1-P_{fc}$
in a typical ``house'' domain using $\beta=1$ and $L=5$. The solid curve is the analytical prediction of \eqref{final1}, while the black dots depict the data from computer simulations.
Also shown using dashed curves are the various contributions for the bulk $U$ (green), the surface area $F$  (yellow), the edges $E_1 +E_2$ (purple), and the corners $C_1+C_2$ (blue).}%
\label{fig:Pout}%
\end{center}
\end{figure}

In order to validate the methodology detailed above, we compare the general formula for the ``house'' example with numerical results obtained through computer simulations. 
Letting $\beta=1$ and $L=5$ for simplicity, we illustrate the accuracy of
our results at high density by plotting the probability of network outage
({i.e.}, $P_{out}=1-P_{fc}$) in Fig.~\ref{fig:Pout}. 
In the figure, the solid curve is the analytical predictions  of \eqref{final1}, while the black dots depict the data from computer simulations. 
We have confirmed our results for other parameters $(\beta,L)$ as well but do not include here for the sake of brevity.

In the simulations, spatial coordinates for $N$ nodes are chosen at random inside a ``house" domain defined by $\V$ as shown in Fig.~\ref{fig:house}. 
The nodes are then paired up whenever a randomly generated number $\wp\in[0,1] < H(r_{i,j})$, where $r_{i,j}$ is the distance between the pair $(i,j)$. 
This guarantees that the links between pairs of nodes are statistically independent.
We store the resulting graph connections in a symmetric adjacency matrix and initiate a depth-first search algorithm to identify the connected components of the graph and whether the graph is fully connected.
Thus, the computational complexity of our algorithm is of order $\mathcal{O}(N \ln N)$.
The process is then repeated in a Monte Carlo fashion and for different values of $N$, thus producing Fig.~\ref{fig:Pout}.

Excellent agreement is achieved at high densities while the approximation is poor at medium to low densities as expected for the reasons described in the previous subsection.
In fact, it is clear that for $\rho<0.3$, the theoretical predictions diverge. 
Also shown using dashed curves are the various contributions for the bulk $U$ (green), the surface area $F$  (yellow), the edges $E_1 +E_2$ (purple), and the corners $C_1+C_2$ (blue).
It becomes clear in Fig.~\ref{fig:Pout} that the corner contribution indeed dominates $P_{fc}$ at high densities (e.g. when $\rho>1$).

\subsection{Example 2: the Half-Cylinder\label{sec:cylinder}}

\begin{figure}
[ptb]
\begin{center}
\includegraphics[width=2.45in]%
{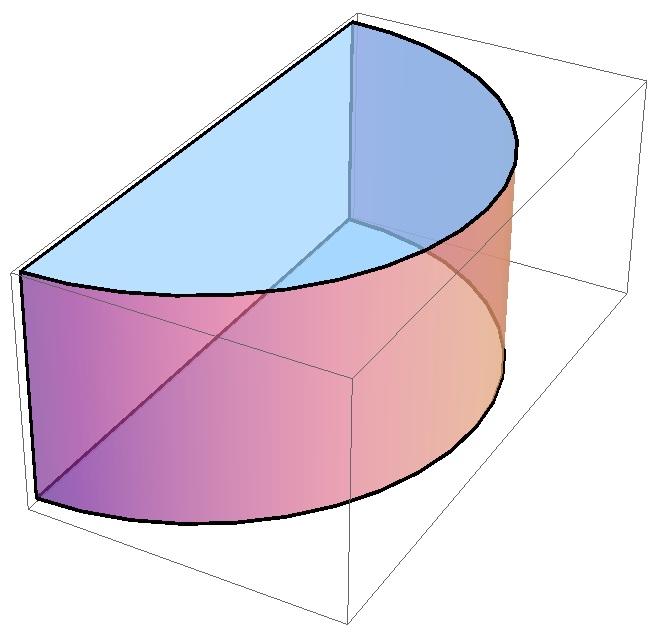}%
\caption{The ``half-cylinder" domain considered for Example 2. Base radius $r$ and height $h$.}%
\label{fig:cylinder}%
\end{center}
\end{figure}

\begin{figure}
[ptb]
\begin{center}
\includegraphics[width=3in
]%
{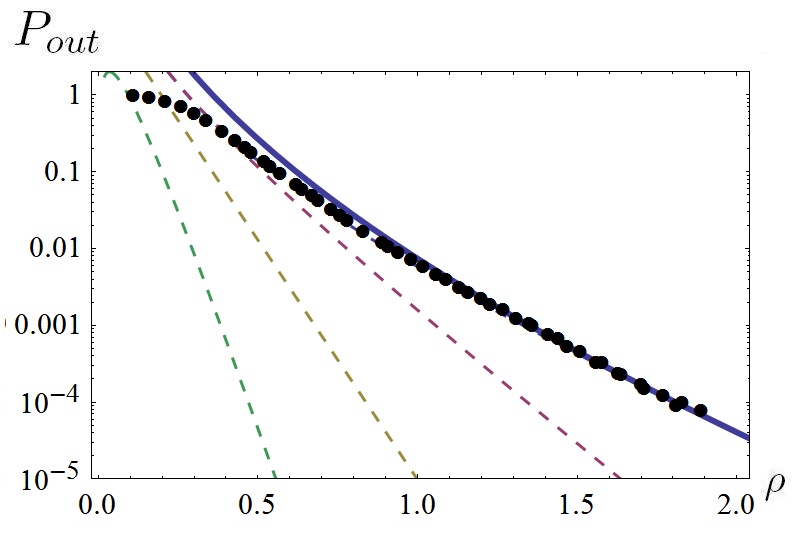}%
\caption{Analytical and numerical results for the network outage probability
in a ``half-cylinder'' domain using $\beta=1$, $r=5$ and $h=4$. The labelling of the curves is identical to Fig.~\ref{fig:Pout}.}%
\label{fig:Pout2}%
\end{center}
\end{figure}

In order to show the versatility of our results, we will adapt and reuse the calculations of Sec.~\ref{sec:general} in order to calculate $P_{fc}$ for another domain shape, namely a ``half-cylinder" domain of base radius $r$ and height $h$ (see Fig.~\ref{fig:cylinder}). Assuming that $\sqrt{\beta} r\gg 1$, and $\sqrt{\beta} h\gg 1$ are sufficient conditions for the validity of our previous approximations.
As we confirm numerically in Fig.~\ref{fig:Pout2}, corrections due to the curved surface of the present domain are of secondary importance at high node densities.
This is particularly true if the local radius of curvature at any point on the surface of $\V$ is much larger than the effective communication range $r_0= \beta^{-1/\eta}$.

Having done all the work in Sec~\ref{sec:general}, it is straight forward to write down 
\es{
P_{fc} \approx 1- \rho( U+ F + E + C),
}
where the various contributions are given by
\es{
U&= \frac{\pi r^2 h}{2} e^{-\frac{(23-\sqrt{2}) \pi^{3/2} \rho}{4 \beta^{3/2}}} \\
F&= \pr{\pi r^2 +2 r h + \pi r h } \frac{2 \beta}{7\pi \rho} e^{-\frac{(23-\sqrt{2}) \pi^{3/2} \rho}{8 \beta^{3/2}}}  \\
E&=  \pr{2\pi r + 4 r +2 h} \frac{16 \beta^2}{49 \pi^2 \rho^2} e^{-\frac{(23-\sqrt{2}) \pi^{3/2} \rho}{16 \beta^{3/2}}} \\
C&= 4\frac{512\beta^{3}}{343\pi^{3}\rho^{3}} e^{-\frac{(23-\sqrt{2}) \pi^{3/2} \rho}{32 \beta^{3/2}}}
.}
The above analytical predictions for the half-cylinder domain are validated through computer simulations and the results is plotted in Fig.~\ref{fig:Pout2}.
It is therefore evident that $P_{fc}$ at high node densities can be easily calculated with the aid of Table \ref{tab:1} for any 3D convex right prism\footnote{Strictly speaking a half-cylinder is not a prism, however it is extremely similar.}.
Furthermore, different fading models (e.g. Rician or Nakagami see \cite{miorandi2008impact,haenggi2008geometric}) can also be accommodated for by performing the calculations outlined in Sec.~\ref{sec:general}.

\section{Extending the Methodology to General Convex Geometries}
Many aspects of the methodology outlined above appear to be loosely related (at best) to the assumption that the bounding polyhedron is a convex right prism.  While it is clear that the choice of a cylindrical coordinate system for analyzing corner contributions is based on the right prism assumption, techniques such as the local equivalence method for considering higher dimensional features deliberately impose a level of abstraction in order to simplify matters.  Thus, one might conjecture that a similar approach could be used to extend the techniques detailed above to more general convex bounding polyhedra.  Here, we provide a small step towards this generalization by presenting an approximation in the spirit of the local equivalence principle that can be used for corners.

A key difference (locally) between a right prism and a general polyhedron is that a corner is created in the former by an intersection of three planes, two of which are normal to the third, whereas a corner can be created in the latter by an intersection of possibly more than three planes without any condition on orientation.  This leads to analytical complications.  However, we may attempt to approximate a general corner contribution by considering the contribution of a \emph{cone} with the same solid angle as the corner in question.  

\begin{figure}
[ptb]
\begin{center}
\includegraphics[width=3in
]%
{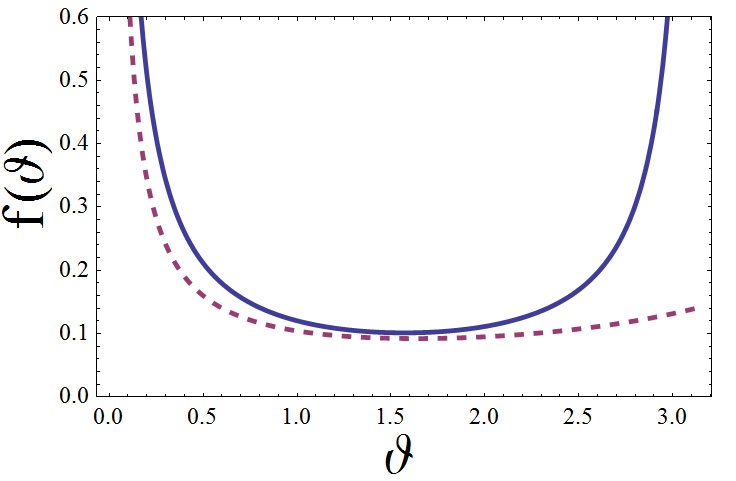}%
\caption{Comparison of the corner contribution (solid line) and the cone contribution (dashed line), normalized by common factors.  For the corner, $f(\vartheta) = \pi^{-1}\csc \vartheta$, whereas for the cone, $f(\vartheta) = 4 \pi^4 \vartheta^{-1} (\vartheta^2 - 6\pi\vartheta + 8 \pi^2)^{-2}$.}
\label{fig:corner_approx}%
\end{center}
\end{figure}

As an illustration of the benefits and deficiencies of this approach, let us consider the approximation of a corner vis-\`a-vis the ``house'' prism example discussed above.  The solid angle of such a corner is the same numerical value as the dihedral angle $\vartheta$, but in units of steradians.  Thus, we wish to approximate the contribution of such a corner to the overall probability of connectivity by substituting a cone of solid angle $\vartheta = 2 \pi (1-\cos \lambda)$ into the analysis, where $2\lambda$ is the apex angle of the cone.  Using spherical coordinates aligned such that the $z$-axis forms an axis of rotation for the cone, we can calculate the contribution in terms of the angle $\vartheta$ to be\footnote{Details are omitted for brevity.}
\begin{equation}
  \int_{\mathcal{V}}e^{-\rho \int_{\mathcal{V}}H (  \mathbf{r}_{12} )
\mathrm{d}\mathbf{r}_{1}  }%
\mathrm{d}\mathbf{r}_{2} = \frac{1024\beta^{3}\pi^4 e^{-\frac{ (  23-\sqrt{2} )  \sqrt{\pi}\rho\vartheta}{16\beta^{3/2}}%
}}{343\rho^{3}\vartheta^2(\vartheta^2 - 6\pi\vartheta + 8\pi^2)^2}.
\label{eq:corner_cone}%
\end{equation}
Comparing with~\eqref{eq:corner}, we immediately note that this expression is of the same order in $\rho$ as was shown for the right-angled corner, which follows the general formula~\eqref{eq:general_formula}.  We also see that the two corner types admit the same order in $\beta$, and thus variations in SNR will be observed similarly for the two cases.  However, the difference occurs in the order of $\vartheta$.  This is to be expected since a cone is a poor approximation for a corner formed by the intersection of three planes.  For more general ``higher order'' corners, however, the conic approximation may be more appropriate.  Nevertheless, by cancelling like terms from both~\eqref{eq:corner} and ~\eqref{eq:corner_cone}, we can graphically compare the two contributions, as shown in Fig.~\ref{fig:corner_approx}.  Note that the approximation is quite good until $\vartheta \rightarrow \pi$, at which point the right-angled expression diverges since we violate some underlying assumptions. 
Namely, at this limit two of the three planes making up the corner become parallel. As a result we may no longer treat this boundary contribution as a corner but rather as an edge. Hence our approximations fail and the contribution plotted in Fig.~\ref{fig:corner_approx} diverges\footnote{See~\cite{Coon2012a}, which treated a general wedge shape in two dimensions.}.
These preliminary results provide encouragement for exploring similar approaches to analyzing general convex polyhedra in the future.

\section{Conclusions\label{sec:conc}}
In this paper, we presented a systematic methodology for analyzing the probability of full connectivity for dense networks confined within convex right prisms. 
We demonstrated the versatility of our approach with two examples, and confirmed our results through extensive numerical simulations.
Moreover, we have suggested avenues that could be explored to extend the theory to more general convex bounding geometries. 
Such extensions may be built upon the use of local approximations to actual geometrical features (e.g., a cone could replace a polyhedral corner), and it is hoped that further developments in this area will lead to a full theory of network connectivity for both convex and nonconvex domains.
One such candidate is the 3D generalization of recent work \cite{6629807} involving the network connectivity through openings such as doors and windows.

\section*{Acknowledgements}
\addcontentsline{toc}{section}{Acknowledgment}
The DIWINE consortium would like to acknowledge the support of the European
Commission partly funding the DIWINE project under Grant Agreement CNET-ICT-318177.
The authors would also like to thank the directors of the Toshiba Telecommunications Research Laboratory for their support.

%%%%%%%%%%%%%%%%%%%%%%

\bibliographystyle{IEEEtran}
\bibliography{connectivity}

\end{document}